\documentclass[aps,prd,preprint,nofootinbib,11pt]{revtex4}
\usepackage{amsfonts}
\usepackage{mathrsfs}
\usepackage{graphicx}
\usepackage{amsmath}
\usepackage{amssymb}
\usepackage{subfigure}
\usepackage{epsfig}
\usepackage{graphicx}
\usepackage{ulem}
\usepackage{color}
\newcommand{\bqa}{\begin{eqnarray}}
\newcommand{\eqa}{\end{eqnarray}}
\newcommand{\beq}{\begin{equation}}
\newcommand{\eeq}{\end{equation}}
\allowdisplaybreaks[1]
\graphicspath{{fig/}{dia/}} \DeclareGraphicsExtensions{.eps}

\begin{document}
\title{\Large QCD sum rule analysis of local meson-meson currents for the $K(1690)$ state\\[7mm]}

\author{Yi-Qi Mu$^{1}$, Peng-Wen Xu$^{1}$, Si-Tong Chen$^{1}$, Yi-Tong Wei$^{1}$, \\Ge-Jia Zhang$^{1}$, and Bing-Dong Wan$^{1,2}$\footnote{wanbd@lnnu.edu.cn} \vspace{+3pt}}

\affiliation{$^1$School of Physics and Electronic Technology, Liaoning Normal University, Dalian 116029, China\\
$^2$ Center for Theoretical and Experimental High Energy Physics, Liaoning Normal University, Dalian 116029, China
}
\author{~\\~\\}

\begin{abstract}
\vspace{0.3cm}
The nature of the recently observed $K(1690)$ state, reported by the COMPASS Collaboration as a candidate strange crypto-exotic meson with $J^P=0^-$, remains unclear. In this work, we investigate whether it can be described by local meson-meson currents within the framework of QCD sum rules. We construct a set of independent local color-singlet meson-meson interpolating currents with $J^P=0^-$ and analyze their QCD sum rules. For the currents that admit reliable Borel windows, the extracted effective mass scales are consistently in the range of $2.0$--$2.3~\mathrm{GeV}$, significantly above the experimental mass of the $K(1690)$. Within the standard pole-plus-continuum framework, none of the selected local currents yields a separately resolved low-lying pole compatible with the COMPASS signal. These results indicate that the selected local color-singlet meson-meson currents do not isolate the $K(1690)$ signal within the present framework, while leaving open the possibility of more extended molecular dynamics and other configurations beyond the present analysis.
\end{abstract}
 \maketitle
\newpage

\section{Introduction}

Hadron spectroscopy provides a fundamental probe of the nonperturbative regime of quantum chromodynamics (QCD)~\cite{Brambilla:2019esw}. While the conventional quark model successfully describes most mesons as quark--antiquark ($q\bar q$) states~\cite{Godfrey:1985xj, Ebert:2009ub}, a growing number of experimental observations indicate the existence of exotic hadrons that cannot be accommodated within this simple picture~\cite{Chen:2016qju, Esposito:2016noz, Ali:2017jda,Wang:2025sic,Liu:2019zoy}. These include compact multiquark configurations, hadronic molecules, and structures induced by kinematical effects~\cite{Guo:2017jvc}. Understanding the internal structure of such states has become a central issue in hadron physics.

Recently, the COMPASS Collaboration reported a high-statistics partial-wave analysis of the reaction $K^- p \to K^- \pi^- \pi^+ p$, in which a resonance structure around $1.7~\mathrm{GeV}$ was observed in the $J^P=0^-$ channel and denoted as $K(1690)$~\cite{COMPASS:2025wkw}. This signal appears as a supernumerary state in the strange-meson spectrum, where quark-model calculations typically predict only two pseudoscalar excitations in this mass region~\cite{Godfrey:1985xj, Ebert:2009ub}. The presence of an additional state therefore points to a nonconventional origin and makes the $K(1690)$ a promising candidate for a strange crypto-exotic meson.

Several theoretical interpretations have been proposed for the $K(1690)$. In particular, compact multiquark configurations have been explored, and QCD sum-rule studies based on tetraquark currents have reported mass predictions compatible with the experimental value~\cite{Zhang:2025fuz}. On the other hand, the hadronic molecular scenario, in which the state is interpreted as a loosely bound system of two mesons, provides an alternative explanation~\cite{Weinstein:1990gu, Tornqvist:1993ng}. Such a picture has been remarkably successful in describing many exotic candidates in the heavy-quark sector~\cite{Guo:2017jvc}. However, its applicability to the light strange sector, where binding effects are expected to be weaker and strong decay channels are abundant, remains highly nontrivial.

From the perspective of exotic configurations, the $K(1690)$ may also be interpreted as a meson-meson molecular system. In such a scenario, a natural quark-level configuration corresponds to a $q\bar{q}$--$q\bar{s}$ structure with flavor content $ud\bar{d}\bar{s}$. As a member of the $K$ meson family, the $K(1690)$ is expected to carry isospin $I=1/2$, which constrains the possible flavor configurations at the quark level. Although different quark rearrangements with the same flavor content are in principle possible, the present work focuses on meson-meson-type configurations corresponding to a $q\bar{q}$--$q\bar{s}$ structure. At the hadron level, these local currents may couple most directly to channels such as pseudoscalar--scalar and vector--axial-vector combinations, schematically represented by $Kf_0/a_0$, $\pi K_0^*$, $K^*K_1$, and related strange light-meson thresholds with the same quantum numbers. This provides a suitable basis for constructing interpolating currents within the QCD sum-rule framework. Nevertheless, it remains unclear whether meson-meson interactions can dynamically generate a low-lying bound state with $J^P=0^-$ at the mass scale of the $K(1690)$. This raises a central question: can meson-meson interactions in the light strange sector generate a bound state at such a low mass scale within a controlled nonperturbative framework?

To the best of our knowledge, QCD sum-rule studies focusing on local meson-meson currents in the $K(1690)$ channel with quark content $ud\bar{d}\bar{s}$ and quantum numbers $J^P=0^-$ remain relatively unexplored.
In this work, we perform a systematic investigation of selected local meson-meson currents in the $K(1690)$ channel within the framework of QCD sum rules~\cite{Shifman:1978bx, Reinders:1984sr, Narison:2002woh}. We construct five independent local color-singlet meson-meson-type interpolating currents with quantum numbers $J^P=0^-$, covering the representative Dirac structures, including $0^- \otimes 0^+$, $0^+ \otimes 0^-$, $1^- \otimes 1^+$, $1^+ \otimes 1^-$, and one independent tensor configuration. We construct the physical $I=1/2$ flavor combination and perform targeted consistency checks of the flavor projection and OPE truncation, while retaining the original $D\leq8$ setup for the central numerical analysis.

Our analysis reveals a systematic pattern: despite the different Dirac structures and quantum configurations, the selected independent meson-meson-type currents lead to effective mass scales around $2~\mathrm{GeV}$ or higher, significantly exceeding the experimental mass of the $K(1690)$. Within the standard pole-plus-continuum treatment, no separately resolved low-lying pole compatible with the COMPASS signal is obtained from these local color-singlet currents. This indicates that the selected local currents are not sufficient to isolate the $K(1690)$ signal in the present QCD sum-rule framework.

This paper is organized as follows. In Sec.~\ref{Formalism}, we present the construction of the interpolating currents and the QCD sum-rule formalism. In Sec.~\ref{Numerical}, we perform the numerical analysis and extract the corresponding Borel-moment effective mass scales. In Sec.~\ref{discussion}, we discuss the implications of our results for the nature of the $K(1690)$. Finally, Sec.~\ref{conclusion} contains our conclusions.

\section{QCD Sum Rule Formalism}\label{Formalism}

QCD sum rules (QCDSR) provide a well-established nonperturbative framework that connects the underlying dynamics of quantum chromodynamics (QCD) with hadronic observables. Following the pioneering work of Shifman, Vainshtein, and Zakharov~\cite{Shifman:1978bx}, the method is formulated through the analysis of two-point correlation functions constructed from interpolating currents that carry the same quantum numbers and quark-gluon content as the hadronic states under study. By matching the QCD representation to the phenomenological description through dispersion relations and quark--hadron duality, one obtains sum rules that allow the extraction of hadronic properties. This approach has been widely and successfully applied to both conventional and exotic hadrons~
\cite{Albuquerque:2013ija,Wang:2013vex,Govaerts:1984hc,Narison:2002woh,Reinders:1984sr,P.Col,Narison:1989aq,Tang:2021zti,Qiao:2014vva,Qiao:2015iea,Tang:2019nwv,Wan:2020oxt,Wan:2022xkx,Zhang:2022obn,Wan:2024dmi,Tang:2024zvf,Li:2024ctd,Zhao:2023imq,Yin:2021cbb,Yang:2020wkh,Wan:2024pet,Wan:2024ykm,Zhang:2024jvv,Tang:2024kmh,Tang:2016pcf,Tang:2015twt,Qiao:2013dda,Qiao:2013raa,Wan:2020fsk,Wan:2025xhf,Chen:2014vha,Azizi:2019xla,Wang:2017sto,Wan:2025bdr,Wan:2025zau,Wan:2025ikc,Wan:2025sae,Tang:2025ept,Ben:2025wqn,Zhang:2025vqg,Zhang:2024ulk,Zhang:2024asb,Zhang:2024ick,Zhang:2023nxl,Wan:2019ake,Chen:2016ymy,Wang:2021qmn,Wan:2022uie,Wan:2021vny,Wan:2023epq,Zhang:2025qmg,Wan:2024fam,Wan:2025fyj,Wan:2026plq,Wan:2026geg}. In the present work, we apply the QCD sum rule framework to the $K(1690)$ system and construct selected local color-singlet meson-meson currents to examine whether they can generate a low-lying structure compatible with the observed signal.

To investigate selected local meson-meson-type configurations in the $K(1690)$ channel with quantum numbers $J^P=0^-$, we construct the following independent color-singlet interpolating currents with quark content $ud\bar{d}\bar{s}$:
\begin{align}
J_A(x) &= i[\bar{d}_a(x) \gamma_5 u_a(x)]\,[\bar{s}_b(x) d_b(x)], \\
J_B(x) &= i[\bar{d}_a(x) u_a(x)]\,[\bar{s}_b(x) \gamma_5 d_b(x)], \\
J_C(x) &= [\bar{d}_a(x) \gamma_\mu u_a(x)]\,[\bar{s}_b(x) \gamma^\mu \gamma_5 d_b(x)], \\
J_D(x) &= [\bar{d}_a(x) \gamma_\mu \gamma_5 u_a(x)]\,[\bar{s}_b(x) \gamma^\mu d_b(x)], \\
J_T(x) &= [\bar{d}_a(x) \sigma_{\mu\nu} u_a(x)]\,[\bar{s}_b(x) \sigma^{\mu\nu} \gamma_5 d_b(x)],
\end{align}
where $a$ and $b$ denote color indices. Each bracket $[\cdots]$ represents a color-singlet bilinear current corresponding to a mesonlike substructure. Therefore, the interpolating currents above can be interpreted as local realizations of meson-meson configurations, such as $0^- \otimes 0^+$, $0^+ \otimes 0^-$, $1^- \otimes 1^+$, $1^+ \otimes 1^-$, and tensor-type combinations.

The labels such as $0^-\otimes0^+$ and $1^-\otimes1^+$ denote the Dirac structures of the local quark bilinears and should not be interpreted as assigning a current to one unique physical two-meson channel, such as a fixed $\pi K_0(700)$ component. A local four-quark current couples to all hadronic states and scattering continua with the same global quantum numbers. In particular, because the pion and kaon are pseudo-Goldstone bosons, the lowest two-body continua in light channels cannot be treated as narrow poles generated by the local current. Therefore, the Borel moment extracted from the local correlator is not expected to reproduce the lightest two-meson threshold directly; rather, it represents an effective scale determined by the spectral weight in the chosen QCD sum-rule window.

We have retained only one tensor current because the two apparent tensor structures are not independent. Using the four-dimensional identity
\begin{equation}
\sigma_{\mu\nu}\gamma_5=\frac{i}{2}\epsilon_{\mu\nu\alpha\beta}\sigma^{\alpha\beta},
\end{equation}
one finds that
\begin{equation}
[\bar{d}_a \sigma_{\mu\nu} u_a][\bar{s}_b \sigma^{\mu\nu}\gamma_5 d_b]
\end{equation}
and
\begin{equation}
[\bar{d}_a \sigma_{\mu\nu}\gamma_5 u_a][\bar{s}_b \sigma^{\mu\nu} d_b]
\end{equation}
are related by an overall convention-dependent sign after relabeling the contracted Lorentz indices. They therefore yield the same diagonal two-point correlation function and should be counted only once.

The tensor current $J_T$ should be regarded as a local operator with total quantum numbers $J^P=0^-$, rather than as a unique product of two fixed physical mesons; it may couple to a superposition of hadronic configurations carrying the same overall quantum numbers.

Since the physical $K(1690)$ belongs to an isospin-$1/2$ strange-meson multiplet, the charged flavor component written above should be combined into a state with definite total isospin. For a generic Dirac structure $X$, we denote
\begin{equation}
J_{+0}^{X}=\eta_X[\bar d\Gamma_X u][\bar s\Gamma'_X d],
\end{equation}
where $\eta_A=\eta_B=i$ and $\eta_C=\eta_D=\eta_T=1$. Thus, only the two currents containing a pseudoscalar bilinear, $J_A$ and $J_B$, are written with the conventional phase factor $i$. The operator $J_{+0}^{X}$ has the flavor structure of a $\pi^+K^0$ component with $I_3=1/2$, but it is not, by itself, an eigenstate of total isospin. We therefore introduce the corresponding $\pi^0K^+$-type component
\begin{equation}
J_{0+}^{X}=\frac{\eta_X}{\sqrt{2}}\left([\bar u\Gamma_X u]-[\bar d\Gamma_X d]\right)[\bar s\Gamma'_X u],
\end{equation}
and construct the $I=1/2$, $I_3=1/2$ current as follows:
\begin{equation}
J_{1/2}^{X}=\sqrt{\frac{1}{3}}J_{0+}^{X}-\sqrt{\frac{2}{3}}J_{+0}^{X}.
\end{equation}
The corresponding correlation function is
\begin{eqnarray}
\Pi_{1/2}^{X} &=& \frac{1}{3}\Pi_{0+,0+}^{X}+\frac{2}{3}\Pi_{+0,+0}^{X} 
-\frac{\sqrt{2}}{3}\left(\Pi_{0+,+0}^{X}+\Pi_{+0,0+}^{X}\right),
\end{eqnarray}
where the last two terms denote the off-diagonal crossed contractions. We evaluate these contractions explicitly. In the isospin limit, $m_u=m_d\equiv m_\ell$ and $\langle\bar uu\rangle=\langle\bar dd\rangle$, the flavor Wick contractions can be organized into a direct two-loop topology, $D_X$, and a one-fermion-loop exchange topology, $E_X$. Including the fermionic Wick signs, one obtains
\begin{equation}
\Pi^X_{+0,+0}=D_X,\qquad
\Pi^X_{0+,0+}=D_X-\frac12 E_X,\qquad
\Pi^X_{0+,+0}=\Pi^X_{+0,0+}=-\frac{1}{\sqrt2}E_X.
\end{equation}
Substituting into Eq.~(12) yields the fully projected correlator.
\begin{equation}
\Pi^X_{1/2}=D_X+\frac12 E_X,
\end{equation}
whereas the orthogonal $I=3/2$ combination gives $\Pi^X_{3/2}=D_X-E_X$. The vanishing of the $I=1/2$--$I=3/2$ mixed correlator provides an algebraic check of the flavor coefficients and Wick signs. We have generated the Wilson coefficients of $E_X$ using the same light-quark propagator, condensate definitions, and vacuum-saturation prescription as in the diagonal OPE, including all Borel-surviving terms up to dimension twelve. We use the complete $I=1/2$ projected correlator as an independent consistency check of the central analysis. In this projected rerun, we impose the isospin-symmetric light-quark input $m_\ell=(m_u+m_d)/2$ and obtain effective scales of approximately $2.04$, $1.99$, $2.33$, and $2.30~\mathrm{GeV}$ for $J_A$--$J_D$, respectively. These values are close to the original $D\leq8$ results and do not yield a low-lying solution compatible with the $K(1690)$. The central masses, Borel windows, and curves quoted in Sec.~\ref{Numerical} are nevertheless retained as those of the initially submitted $D\leq8$ numerical analysis for direct comparison; the projected calculation is not used to redefine Table~\ref{mass}.

It should be emphasized that these local currents do not correspond to pure molecular states with spatial separation, but instead serve as effective operators that couple to hadronic configurations with the same quantum numbers. In addition, the present operator basis is restricted to selected local color-singlet--color-singlet meson-meson currents; color-octet--color-octet operators and more general linear combinations lie beyond the scope of this work. Therefore, the extracted results should be interpreted as constraints on the selected local color-singlet meson-meson-type configurations, without addressing all possible molecular or non-molecular structures. They should be regarded as probes of meson-meson-type structures rather than strict bound-state wave functions.

Based on the above interpolating currents, we construct the corresponding two-point correlation functions,
\begin{equation}
\Pi(q^2) = i \int d^4 x \, e^{i q \cdot x} \langle 0 | T [ J(x) J^\dagger(0) ] | 0 \rangle \; ,
\end{equation}
which serve as the fundamental objects in the QCD sum rule approach and encode the QCD dynamics of the hadronic system under consideration.

On the QCD side, the correlation function admits a dispersion representation, which follows from its analytic properties in the complex $q^2$ plane. Within the framework of the operator product expansion, it can be further expressed as
\begin{equation}
\Pi^{\text{OPE}}(q^2) = \int_{s_{\text{min}}}^{\infty} ds \, \frac{\rho^{\text{OPE}}(s)}{s - q^2} \; ,
\end{equation}
where $s_{\text{min}}$ denotes the kinematic threshold, given by the squared sum of the relevant quark masses, and $\rho^{\text{OPE}}(s)$ encodes the quark--gluon dynamics, including both perturbative contributions and nonperturbative effects from QCD vacuum condensates. In practical calculations, the dispersive spectral density is written as a sum of perturbative, quark-condensate, gluon-condensate, mixed-condensate, and higher-dimensional terms. In addition, the coordinate-space OPE contains non-dispersive terms. For the central effective-mass analysis below, the OPE is truncated at dimension eight. To test this truncation explicitly, we also derive the terms generated at dimensions nine through twelve using the same propagator and vacuum-saturation prescription. For the currents considered here, the dimension-nine and dimension-ten extensions are non-dispersive, whereas the generated dimension-eleven and dimension-twelve terms reduce to local contact terms and are annihilated by the Borel transform. Their analytic expressions and numerical fractions relative to the $D\leq8$ Borel moment are reported as convergence diagnostics, but they are not included in the quoted central effective-mass analysis.

On the phenomenological side, the same correlation function is described in terms of hadronic degrees of freedom, with the lowest-lying state explicitly separated from the higher resonances and the continuum:
\begin{equation}
\Pi^{\text{phen}}(q^2) = \frac{\lambda^2}{M^2 - q^2} + \int_{s_0}^{\infty} ds \, \frac{\rho(s)}{s - q^2} \; ,
\end{equation}
where the first term represents the pole contribution in the adopted ansatz, whereas the second term accounts for higher resonances and the continuum. Here, $M$ denotes the mass parameter associated with the pole term, $\lambda$ is the pole residue describing the coupling strength of the interpolating current to this contribution, and $s_0$ is the continuum threshold separating the pole term from higher excited states and continuum contributions.

We emphasize that Eq.~(17) is the standard narrow-pole-plus-continuum parametrization used in QCD sum rules. It does not assume that the entire physical spectral density is a delta function; rather, only the lowest pole term is approximated as a narrow contribution, while higher resonances and continuum states are modeled through quark--hadron duality above $s_0$. For light multiquark channels and local color-singlet meson-meson currents, however, this parametrization has an important limitation: the currents can couple strongly to two-hadron continua, and a shallow or broad near-threshold structure may not be resolved as an isolated pole. Therefore, the quantity extracted from the ratio of Borel moments should be interpreted, in the present context, as a Borel-moment effective mass scale unless isolated pole dominance is clearly established.

To relate the QCD representation to the phenomenological one, we invoke the quark--hadron duality hypothesis, under which the continuum contribution is approximated by the OPE spectral density above a threshold $s_0$:
\begin{equation}
\int_{s_{\text{min}}}^{\infty} ds \, \frac{\rho^{\text{OPE}}(s)}{s - q^2}
=
\frac{\lambda^2}{M^2 - q^2}
+
\int_{s_0}^{\infty} ds \, \frac{\rho^{\text{OPE}}(s)}{s - q^2} \; ,
\end{equation}
where the second term on the right-hand side represents contributions from higher resonances and the continuum, which are approximated by the OPE spectral density above the threshold $s_0$.

With this approximation, the contributions from higher resonances and the continuum can be effectively removed by subtracting the spectral integral above $s_0$. One then obtains
\begin{equation}
\frac{\lambda^2}{M^2 - q^2}
=
\int_{s_{\text{min}}}^{s_0} ds \, \frac{\rho^{\text{OPE}}(s)}{s - q^2} \; .
\end{equation}

To further suppress contributions from higher states and improve the convergence of the operator product expansion, a Borel transformation is applied to both sides of the sum rule. Because the OPE also contains non-dispersive terms, the continuum-subtracted Borel sum rule is written as
\begin{equation}
\lambda^2 e^{-M^2/M_B^2}
=
\int_{s_{\text{min}}}^{s_0} ds \, \rho^{\text{OPE}}(s) e^{-s/M_B^2}
+\Pi_{\rm nd}(M_B^2) \; ,
\end{equation}
where $\Pi_{\rm nd}$ collects the Borel-surviving, non-dispersive condensate terms included in the adopted $D\leq8$ mass sum rule. The analytically derived $D=9$ and $D=10$ non-dispersive terms are used only to estimate the magnitude of the omitted higher-dimensional tail, whereas the generated $D=11$ and $D=12$ terms are contact terms and therefore do not contribute after the Borel transform.

For convenience, we introduce the Borel moment
\begin{equation}
L_0(s_0,M_B^2) = \int_{s_{\text{min}}}^{s_0} ds \, \rho^{\text{OPE}}(s) e^{-s/M_B^2}+\Pi_{\rm nd}(M_B^2) \; ,
\end{equation}
which represents the continuum-subtracted, Borel-weighted OPE moment below $s_0$, including the non-dispersive contribution.

To extract the effective mass scale, it is useful to consider the derivative of the sum rule with respect to $-1/M_B^2$. This procedure enhances the contribution from higher-energy regions and makes the mass dependence explicit:
\begin{equation}
\frac{\partial}{\partial(-1/M_B^2)} L_0(s_0,M_B^2)
=
\int_{s_{\text{min}}}^{s_0} ds \, s \, \rho^{\text{OPE}}(s) e^{-s/M_B^2}
+\frac{\partial \Pi_{\rm nd}}{\partial(-1/M_B^2)} \; .
\end{equation}

We therefore define
\begin{equation}
L_1(s_0,M_B^2)
=
\frac{\partial}{\partial(-1/M_B^2)} L_0(s_0,M_B^2) \; .
\end{equation}

Taking the ratio of the two sum rules eliminates the unknown pole residue $\lambda$ and allows the Borel-moment effective mass to be determined within the adopted pole-plus-continuum ansatz:
\begin{equation}
M_{\rm eff}^2(s_0,M_B^2)
=
\frac{L_1(s_0,M_B^2)}{L_0(s_0,M_B^2)} \; ,
\end{equation}
It is the ratio of the first and zeroth full Borel OPE moments used in the $D\leq8$ analysis, including both dispersive and non-dispersive contributions. In the purely dispersive limit, it reduces to a Borel-weighted average over the spectral distribution below $s_0$. This quantity can be identified with a hadron mass only when the corresponding current is dominated by a sufficiently isolated low-lying pole; in the present analysis, we refer to it as the Borel-moment effective mass scale $M_{\rm eff}$.

Accordingly, the Borel-moment effective mass scale can be written as
\begin{equation}
M_{\rm eff}(s_0,M_B^2)
=
\sqrt{\frac{L_1(s_0,M_B^2)}{L_0(s_0,M_B^2)}} \; .
\end{equation}

\section{Numerical Analysis}\label{Numerical}

To ensure direct reproducibility of the originally submitted numerical analysis, the central mass values quoted below retain the original input set,
$m_u=2.3~\mathrm{MeV}$,
$m_d=6.4~\mathrm{MeV}$,
$m_s=95~\mathrm{MeV}$,
$\langle \bar q q\rangle=-(0.23~\mathrm{GeV})^3$,
$\langle \bar s s\rangle=0.8\langle\bar q q\rangle$,
$\langle \bar q g_s\sigma\cdot G q\rangle=m_0^2\langle\bar q q\rangle$,
$\langle \bar s g_s\sigma\cdot G s\rangle=m_0^2\langle\bar s s\rangle$,
$\langle g_s^2G^2\rangle=0.88~\mathrm{GeV}^4$, and $m_0^2=0.8~\mathrm{GeV}^2$, together with the parameter variations used in the original uncertainty estimate. Because the referee correctly noted that the quark masses and condensates in this original set are conventionally quoted at different reference scales, we additionally perform a scale-consistency check in which the light-quark masses are evolved to $\mu\simeq1~\mathrm{GeV}$ and all scale-dependent inputs are evaluated at that common hadronic scale. The resulting shifts in the effective mass scales are small compared with the conservative uncertainties quoted in Table~\ref{mass}; therefore, the original central values are retained for direct comparison, while the common-scale rerun is used as a robustness check rather than as a redefinition of the numerical analysis.

Furthermore, within the QCD sum rule framework, two auxiliary parameters are introduced, namely the continuum threshold $s_0$ and the Borel parameter $M_B^2$. Their working regions are determined following the standard procedures described in Refs.~\cite{Shifman:1978bx,Reinders:1984sr,P.Col}, which are based on two well-established criteria.

The first criterion concerns the convergence of the operator product expansion (OPE). In practice, the relative contribution of each higher-dimensional term is compared with the total OPE, and an appropriate Borel window is chosen such that the truncated OPE remains under control. It should be emphasized that higher-dimensional condensates correspond to contributions with lower powers of $s$ in the spectral density. As a result, OPE convergence is assessed by evaluating the relative weight of the low-power terms in $s$ and ensuring that their contributions are sufficiently suppressed. Physically, this requirement ensures that the nonperturbative contributions encoded in higher-dimensional condensates do not dominate the sum rule, so that the truncated OPE remains a reliable approximation.

The second criterion is related to the pole contribution (PC). For multiquark sum rules in the present analysis, the pole contribution is required to be larger than $50\%$ of the total, ensuring that the pole term in the adopted ansatz is sufficiently important. This condition is used to suppress contamination from higher resonances and the continuum, although it does not by itself prove that a narrow physical ground state has been isolated.

These two conditions can be expressed as
\begin{eqnarray}
  R^{OPE} = \left| \frac{L_{0}^{c_0}(s_0, M_B^2)}{L_{0}(s_0, M_B^2)} \right| \le 10\%\, ,
\end{eqnarray}
\begin{eqnarray}
  R^{PC} = \frac{L_{0}(s_0, M_B^2)}{L_{0}(\infty, M_B^2)}\ge 50\% \; , \label{RatioPC}
\end{eqnarray}
where the superscript $c_0$ denotes the contribution to the Borel-transformed sum rule from the term in the spectral density $\rho^{\mathrm{OPE}}(s)$ proportional to $s^0$. This criterion is retained to ensure continuity with the original $D\leq8$ numerical analysis. In addition, we explicitly inspect the condensate-by-condensate hierarchy over the full adopted Borel windows and quote the sizes of the analytically derived $D=9$--$12$ terms relative to the $D\leq8$ moment. This provides an independent check that the omitted higher-dimensional tail remains numerically small.

To determine a suitable value of the continuum threshold $s_0$, we adopt a procedure similar to those used in Refs.~\cite{Qiao:2013dda,Tang:2016pcf,Qiao:2013raa}. The goal is to identify a region in which the extracted effective mass exhibits a controlled residual dependence on the Borel parameter $M_B^2$. Within this working window, the dependence of the mass on $M_B^2$ should remain limited. This reflects the requirement that a physically meaningful sum rule should not be dominated by the auxiliary Borel parameter, thereby indicating a proper balance between OPE convergence and pole dominance. In practical calculations, $\sqrt{s_0}$ is varied by $\pm 0.1$ GeV to estimate its uncertainty, following Refs.~\cite{Wan:2020oxt,Wan:2020fsk}.

\begin{figure}
\includegraphics[width=6.8cm]{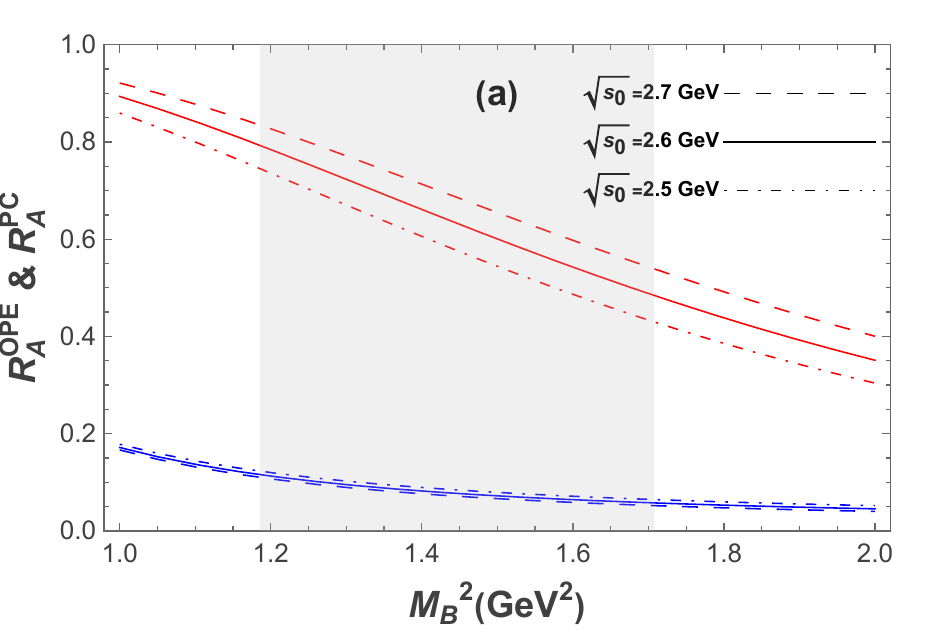}
\includegraphics[width=6.8cm]{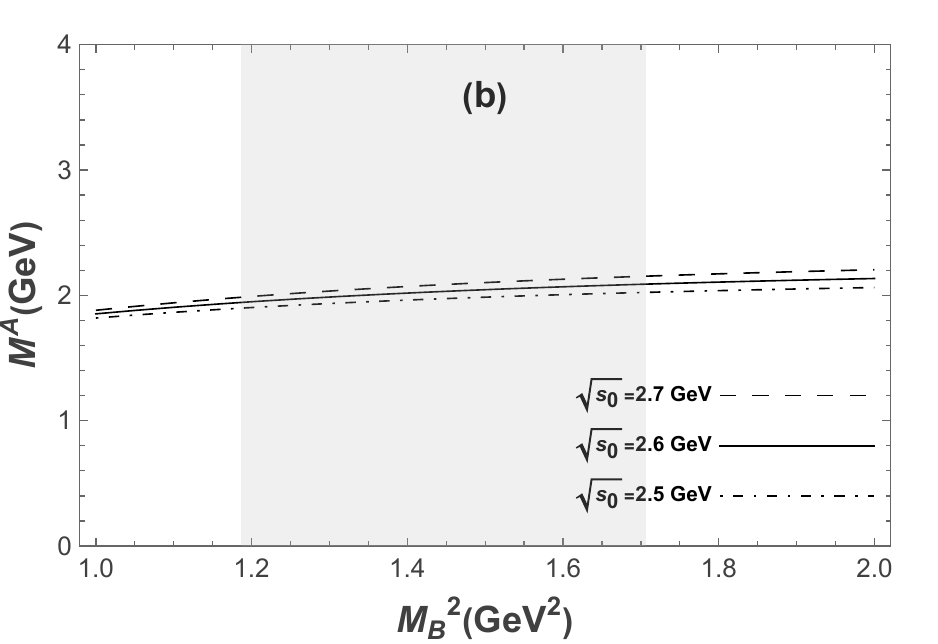}
\caption{ (a) The ratios $R^{OPE}_{A}$ and $R^{PC}_{A}$ as functions of the Borel parameter $M_B^2$ for different values of $\sqrt{s_0}$. The blue curves represent $R^{OPE}_{A}$, the red curves represent $R^{PC}_{A}$, and the different line styles correspond to different values of $\sqrt{s_0}$. The shaded band indicates the Borel window adopted in the numerical analysis. (b) The effective mass $M^{A}$ as a function of the Borel parameter $M_B^2$ for different values of $\sqrt{s_0}$.} \label{figA}
\end{figure}

\begin{figure}
\includegraphics[width=6.8cm]{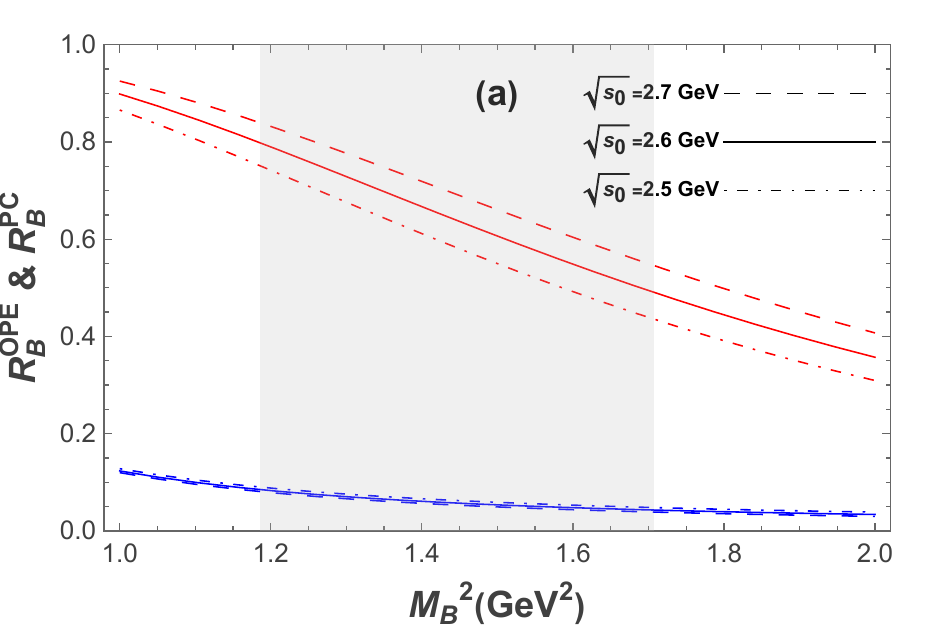}
\includegraphics[width=6.8cm]{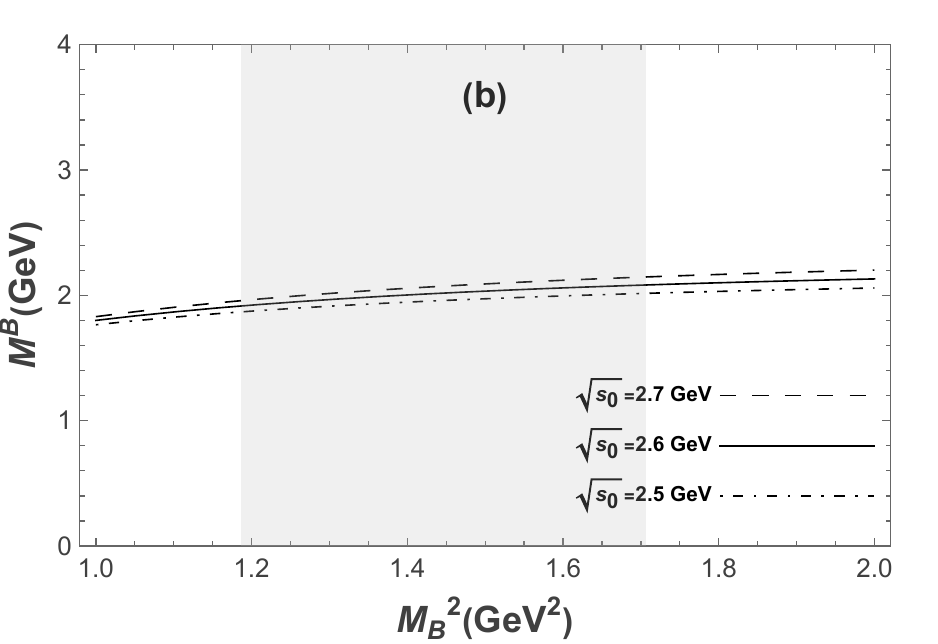}
\caption{ (a) The ratios $R^{OPE}_{B}$ and $R^{PC}_{B}$ as functions of the Borel parameter $M_B^2$ for different values of $\sqrt{s_0}$. The blue curves represent $R^{OPE}_{B}$, whereas the red curves represent $R^{PC}_{B}$; different line styles correspond to different values of $\sqrt{s_0}$. The shaded band indicates the Borel window adopted in the numerical analysis. (b) The effective mass $M^{B}$ as a function of the Borel parameter $M_B^2$.} \label{figB}
\end{figure}

\begin{figure}
\includegraphics[width=6.8cm]{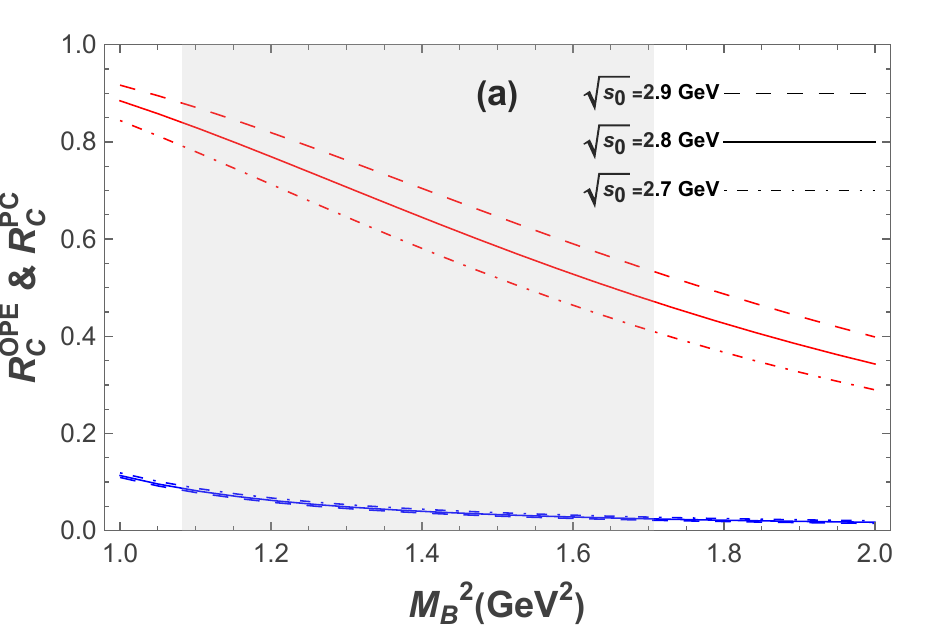}
\includegraphics[width=6.8cm]{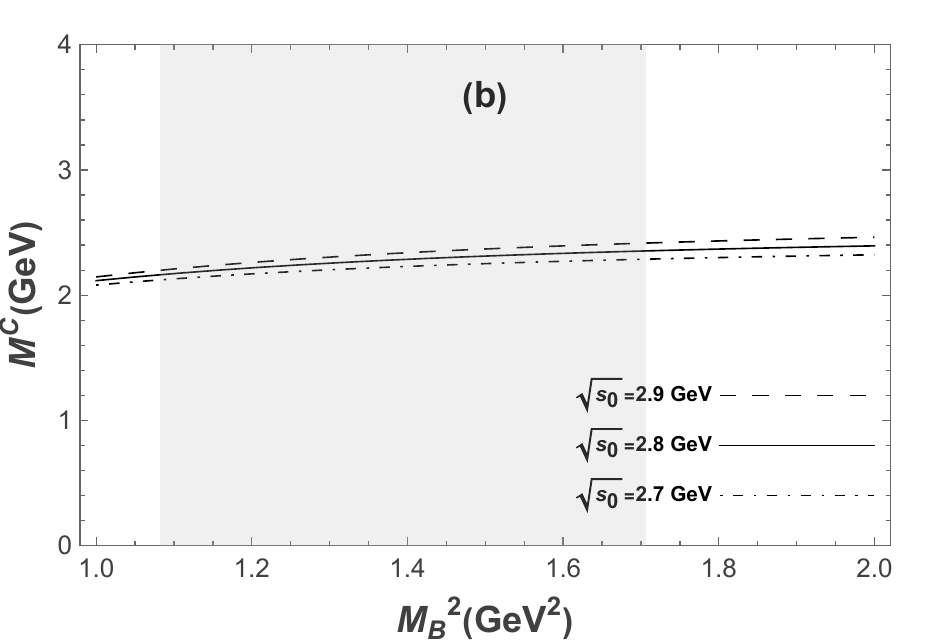}
\caption{ (a) The ratios $R^{OPE}_{C}$ and $R^{PC}_{C}$ as functions of the Borel parameter $M_B^2$ for different values of $\sqrt{s_0}$. The blue curves represent $R^{OPE}_{C}$, the red curves represent $R^{PC}_{C}$, and the different line styles correspond to different values of $\sqrt{s_0}$. The shaded band indicates the Borel window adopted in the numerical analysis. (b) The effective mass $M^{C}$ as a function of the Borel parameter $M_B^2$.} \label{figC}
\end{figure}

\begin{figure}
\includegraphics[width=6.8cm]{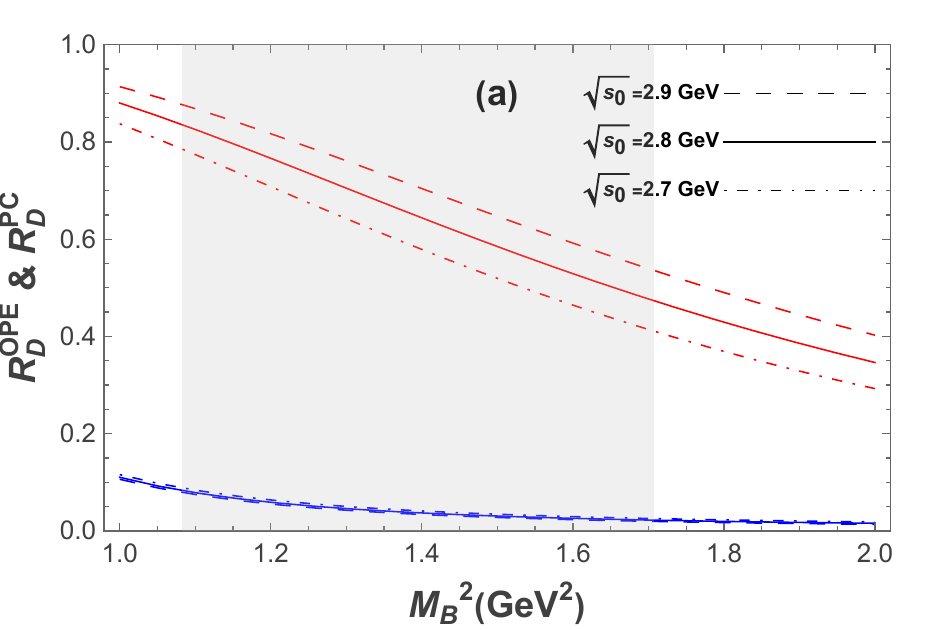}
\includegraphics[width=6.8cm]{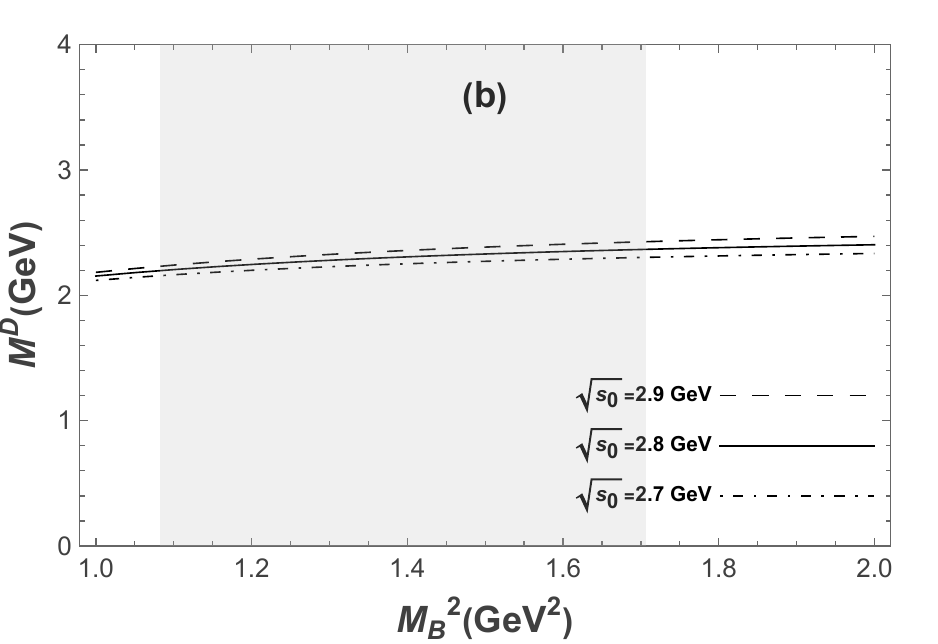}
\caption{ (a) The ratios $R^{OPE}_{D}$ and $R^{PC}_{D}$ as functions of the Borel parameter $M_B^2$ for different values of $\sqrt{s_0}$. The blue curves represent $R^{OPE}_{D}$, and the red curves represent $R^{PC}_{D}$; different line styles correspond to different values of $\sqrt{s_0}$. The shaded band indicates the Borel window adopted in the numerical analysis. (b) The effective mass $M^{D}$ as a function of the Borel parameter $M_B^2$.} \label{figD}
\end{figure}

With the above preparation, the effective mass scales associated with the local meson-meson currents can be evaluated numerically. For direct comparison with the initially submitted analysis, we retain its $D\leq8$ numerical setup and quoted Borel intervals: $1.2$--$1.7~\mathrm{GeV}^2$ for $J_A$ and $J_B$, $1.1$--$1.7~\mathrm{GeV}^2$ for $J_C$, and $1.1$--$1.7~\mathrm{GeV}^2$ for $J_D$. The isospin-projected calculation and the $D=9$--$12$ terms are used as additional consistency and truncation checks and are not used to redefine these central numerical results. For the $J_A$ state, the ratios $R^{OPE}_{A}$ and $R^{PC}_{A}$ are presented as functions of the Borel parameter $M_B^2$ in Fig. \ref{figA}(a) for different values of $\sqrt{s_0}$, i.e., $2.5$, $2.6$, and $2.7$ GeV. The dependence of $M^{A}$ on the Borel parameter $M_B^2$ is displayed in Fig. \ref{figA}(b). We note that the effective mass curves in Figs.~\ref{figA}(b)--\ref{figD}(b) do not form perfectly flat plateaus, but instead show a residual monotonic dependence on $M_B^2$. We therefore interpret the extracted quantities as effective mass scales rather than sharply resolved pole masses. The residual Borel dependence of the extracted effective mass remains controlled in the adopted working region, reflecting a balance between OPE convergence and pole dominance. The adopted Borel window is $1.20 \le M_B^2 \le 1.70\; \text{GeV}^2$, and the effective mass scale $M^{A}$ is
\begin{eqnarray}
M^{A} &=& (2.03\pm 0.13)\; \text{GeV}.
\end{eqnarray}
This indicates that the current $J_A$ yields an admissible sum rule in the adopted working region. Over this window, the original curve has a maximum slope of approximately $0.40~\mathrm{GeV}/\mathrm{GeV}^2$, and the endpoint drift is about $6.9\%$. This residual dependence is included in the conservative uncertainty and is not interpreted as a perfectly flat pole plateau.

For the $J_B$ state, similar OPE-convergence and pole-dominance behaviors can be observed in Fig.~\ref{figB}. An admissible working region is obtained in the range $1.20 \le M_B^2 \le 1.70\; \text{GeV}^2$ with $\sqrt{s_0} = 2.6 \pm 0.1$ GeV. The extracted effective mass is
\begin{eqnarray}
M^{B} &=& (2.01\pm 0.14)\; \text{GeV}.
\end{eqnarray}
The current $J_B$ leads to a similar working region. For the original curve, the maximum slope is approximately $0.48~\mathrm{GeV}/\mathrm{GeV}^2$, and the endpoint drift is about $7.9\%$. This residual dependence is treated as part of the numerical uncertainty rather than as evidence of a perfectly flat plateau.

For the $J_C$ state, the corresponding results are shown in Fig.~\ref{figC}. Following the original numerical setup, we retain the Borel window $1.10 \le M_B^2 \le 1.70\; \text{GeV}^2$ with $\sqrt{s_0} = 2.8 \pm 0.1$ GeV. The extracted effective mass scale is
\begin{eqnarray}
M^{C} &=& (2.25\pm 0.17)\; \text{GeV}.
\end{eqnarray}
For the $J_C$ curve over $1.1$--$1.7~\mathrm{GeV}^2$, the maximum slope is approximately $0.49~\mathrm{GeV}/\mathrm{GeV}^2$, and the endpoint drift is about $7.9\%$. This residual dependence is included in the conservative uncertainty quoted for this channel.

For the $J_D$ state, as shown in Fig.~\ref{figD}, we likewise retain the adopted working region $1.10 \le M_B^2 \le 1.70\; \text{GeV}^2$ with $\sqrt{s_0} = 2.8 \pm 0.1$ GeV, yielding
\begin{eqnarray}
M^{D} &=& (2.29\pm 0.13)\; \text{GeV}.
\end{eqnarray}
For the original $J_D$ curve, the maximum slope is approximately $0.44~\mathrm{GeV}/\mathrm{GeV}^2$, and the endpoint drift is about $7.0\%$. The residual Borel dependence is likewise included in the quoted conservative uncertainty.

The effective mass scales extracted from $J_C$ and $J_D$ are numerically close. We do not regard this as an independent cross-check based on unrelated currents, because $J_C$ and $J_D$ are related as chiral-partner structures and differ mainly through chirally odd contributions in the OPE. The same caution applies to the near degeneracy of $J_A$ and $J_B$.

The uncertainties in the results arise predominantly from imprecisions in the quark masses, vacuum condensates, and the continuum threshold parameter $\sqrt{s_0}$. In practice, we vary these inputs within their quoted ranges and combine the resulting shifts in quadrature. We also monitor the residual dependence on the Borel window and find that it remains smaller than the quoted total uncertainty for the admissible currents. For ease of reference, the Borel parameters, continuum thresholds, and effective mass scales are tabulated in Table~\ref{mass}.

For the tensor interpolating current $J_T$, no admissible Borel window satisfying both the OPE convergence and pole-dominance criteria is found for any reasonable choices of $M_B^2$ and $\sqrt{s_0}$. More specifically, in the low-$M_B^2$ region, where the pole contribution could be relatively enhanced, the OPE convergence deteriorates rapidly because the higher-dimensional condensate terms become numerically important. Increasing $M_B^2$ improves the OPE behavior, but the pole contribution then drops below the required threshold, and the extracted effective mass no longer exhibits controlled Borel behavior. This indicates that the independent tensor current does not resolve a low-lying pole within the present QCD sum rule framework.

From the above results, a clear pattern emerges: all currents that lead to admissible sum rules yield effective mass scales significantly higher than the experimental value, whereas the remaining tensor current does not produce a reliable working window.

\begin{table}
\begin{center}
\renewcommand\tabcolsep{10pt}
\caption{The continuum thresholds, Borel parameters, and effective mass scales associated with the local meson--meson currents.}\label{mass}
\begin{tabular}{cccc}\hline\hline
Current   & $\sqrt{s_0}\;(\text{GeV})$         &$M_B^2\;(\text{GeV}^2)$     &$M^X\;(\text{GeV})$       \\ \hline
$A$        & $2.6\pm0.1$                             &$1.20-1.70$                              &$2.03\pm0.13$         \\
$B$        & $2.6\pm0.1$                             &$1.20-1.70$                               &$2.01\pm0.14$          \\
$C$        & $2.8\pm0.1$                             &$1.10-1.70$                               &$2.25\pm0.17$           \\
$D$        & $2.8\pm0.1$                             &$1.10-1.70$                               &$2.29\pm0.13$           \\
$T$        & $\cdots$                                    &$\cdots$                                &$\cdots$         \\\hline
 \hline
\end{tabular}
\end{center}
\end{table}

As an additional diagnostic, Table~\ref{ope} presents the dimension-by-dimension composition of the $D\leq8$ OPE at representative points. This diagnostic does not replace the original $R^{\rm OPE}$ criterion or modify the Borel windows used in the central mass analysis. The separately derived $D=9$--$12$ terms address a different question: whether the original truncation at $D=8$ is numerically reasonable. Their analytic expressions are provided in Appendix~\ref{app:rho}, and their relative sizes with respect to the retained $D\leq8$ moment are summarized in Table~\ref{highD}.

\begin{table}
\begin{center}
\renewcommand\tabcolsep{5pt}
\caption{Relative contributions of the retained condensate dimensions in the $D\leq8$ OPE used in the central numerical analysis, evaluated at representative points within the adopted Borel windows and defined as $|L_0^{(D)}/L_0^{(D\leq8)}|\times100\%$.}\label{ope}
\resizebox{\columnwidth}{!}{
\begin{tabular}{ccccccccccc}\hline\hline
Current & $\sqrt{s_0}$ & $M_B^2$ & $D=0$ & $D=3$ & $D=4$ & $D=5$ & $D=6$ & $D=7$ & $D=8$ \\ \hline
$A$ & 2.6 & 1.45 & 50\% & 18\% & 59\% & 7\% & 9\% & 2\% & 5\% \\
$B$ & 2.6 & 1.45 & 46\% & 5\% & 55\% & 3\% & 9\% & 1\% & 5\% \\
$C$ & 2.8 & 1.40 & 109\% & 23\% & $<0.1\%$ & 9\% & 9\% & 1\% & 5\% \\
$D$ & 2.8 & 1.40 & 107\% & $<0.1\%$ & $<0.1\%$ & 2\% & 9\% & 1\% & 5\% \\ \hline
\end{tabular}
}
\end{center}
\end{table}

For $J_C$ and $J_D$, the perturbative contribution can exceed $100\%$ because condensate terms with opposite signs partially cancel the perturbative contribution; this reflects a cancellation effect rather than a normalization inconsistency. Across the adopted windows, the $D=8$ contribution ranges approximately from $4.1$--$8.0\%$, $3.9$--$7.4\%$, $3.2$--$10.8\%$, and $3.1$--$11.1\%$ for $J_A$--$J_D$, respectively. Thus, for the corrected $J_C$ window of $1.1$--$1.7~\mathrm{GeV}^2$, the highest retained term remains at approximately the ten-percent level. The central Borel windows and the original $R^{\rm OPE}$ and $R^{\rm PC}$ curves are retained unchanged. The purpose of the new higher-dimensional calculation is more limited: it directly assesses the size of the first terms omitted in the original $D=8$ truncation.

\begin{table}
\begin{center}
\renewcommand\tabcolsep{6pt}
\caption{Diagnostic size of the analytically derived $D=9$--$12$ terms over the full adopted $D\leq8$ Borel windows. The entries represent the ranges of $|L_0^{(D)}/L_0^{(D\leq8)}|\times100\%$. These terms are not included in the quoted central mass extraction.}\label{highD}
\begin{tabular}{ccccc}\hline\hline
Current & $D=9$ & $D=10$ & $D=11$ & $D=12$ \\ \hline
$A$ & 0.026--0.071\% & 0.51--1.37\% & 0 & 0 \\
$B$ & 0.014--0.038\% & 0.43--1.13\% & 0 & 0 \\
$C$ & 0.31--1.61\% & 0.51--2.66\% & 0 & 0 \\
$D$ & 0.11--0.60\% & 0.43--2.40\% & 0 & 0 \\ \hline
\end{tabular}
\end{center}
\end{table}

The explicit $D=9$--$12$ calculation therefore serves only as a diagnostic of the truncation. The first omitted Borel-surviving terms are small compared with the $D\leq8$ moment, while the generated $D=11$ and $D=12$ contributions vanish after the Borel transformation. This supports retaining the original $D=8$ truncation in the central effective-mass analysis without changing the numerical setup used in the initially submitted manuscript.

With the current convention in Eqs.~(1)--(5), in which only $J_A$ and $J_B$ carry the phase factor $i$, the extracted pole residues satisfy $\lambda^2>0$ throughout all adopted Borel windows. At the central Borel points, we obtain $\lambda^2=(3.55,\,3.77,\,17.1,\,19.7)\times10^{-5}~\mathrm{GeV}^{10}$ for $J_A$, $J_B$, $J_C$, and $J_D$, respectively.

To assess the sensitivity to the gluon-condensate input, we repeated the extraction using $\langle g_s^2G^2\rangle=0.47~\mathrm{GeV}^4$ in addition to the central value $0.88~\mathrm{GeV}^4$. The induced shifts in the effective mass scales are at the level of $5$--$50~\mathrm{MeV}$, well below the quoted total uncertainties. This indicates that the results are not sensitive to this input within the present precision.

The individual sources of uncertainty are summarized in Table~\ref{errbudget}. The total uncertainty for each current is obtained by combining the individual contributions in quadrature. The dominant contributions arise from the continuum threshold $\sqrt{s_0}$ and the residual Borel-window dependence, while variations in the quark condensate, gluon condensate, mixed condensate, and strange-quark mass induce subleading shifts.

\begin{table}
\begin{center}
\renewcommand\tabcolsep{6pt}
\caption{Individual contributions to the uncertainty in the effective mass scale (in GeV) for each current.}\label{errbudget}
\begin{tabular}{ccccc}\hline\hline
Source & $J_A$ & $J_B$ & $J_C$ & $J_D$ \\ \hline
$\sqrt{s_0}\pm0.1$ & $\pm0.06$ & $\pm0.06$ & $\pm0.06$ & $\pm0.05$ \\
$\langle\bar qq\rangle\pm30\%$ & $\pm0.005$ & $<0.001$ & $\pm0.008$ & $\pm0.008$ \\
$\langle g_s^2G^2\rangle\pm0.25$ & $\pm0.024$ & $\pm0.027$ & $<0.001$ & $\pm0.011$ \\
$m_0^2\pm25\%$ & $\pm0.001$ & $\pm0.004$ & $\pm0.011$ & $\pm0.005$ \\
$m_s\pm6~\mathrm{MeV}$ & $\pm0.001$ & $<0.001$ & $\pm0.001$ & $\pm0.002$ \\
Borel residual & $+0.06/-0.08$ & $+0.06/-0.10$ & $+0.06/-0.10$ & $+0.05/-0.08$ \\ \hline
\end{tabular}
\end{center}
\end{table}

\section{Discussion}\label{discussion}

In this work, we performed a QCD sum rule analysis of the $K(1690)$ channel using selected independent local color-singlet meson-meson-type interpolating currents with the same quantum numbers.

Our results exhibit a clear and consistent pattern. For the currents $J_A$, $J_B$, $J_C$, and $J_D$, the quoted central mass analysis retains the original $D\leq8$ numerical setup. The full isospin projection and the explicit $D=9$--$12$ calculation are employed as additional consistency and truncation checks. However, all of these currents lead to effective mass scales in the range of approximately $2.0$--$2.3~\mathrm{GeV}$, which is significantly higher than the experimental mass of the $K(1690)$. By contrast, for the tensor current $J_T$, the OPE convergence deteriorates rapidly, and no reliable Borel window can be identified, preventing the extraction of a meaningful mass scale.

This systematic behavior indicates that the selected local meson-meson interpolating currents do not isolate the observed mass of the $K(1690)$ in the present single-current analysis. The near degeneracy between the pairs $J_A/J_B$ and $J_C/J_D$ should not be interpreted as independent confirmation from unrelated currents, since these pairs are connected by chiral transformations and differ mainly in chirally odd OPE contributions. A useful comparison is provided by the compact tetraquark sum rule analysis of Ref.~\cite{Zhang:2025fuz}, which reported a mass compatible with the observed signal using diquark--antidiquark currents. The contrast between that result and the present one indicates that different current choices may probe different dominant couplings, even though they share the same overall flavor quantum numbers.

From a physical perspective, the residual Borel dependence observed for the admissible currents indicates that the extracted quantities should be interpreted as Borel-moment effective mass scales within the adopted pole-plus-continuum ansatz. Therefore, the discrepancy between these effective mass scales and the experimental value should not be overinterpreted as excluding molecular dynamics, but rather as evidence that the selected local currents do not resolve a low-lying pole in the present framework.

This interpretation is especially relevant for local color-singlet--color-singlet currents. Such operators naturally couple not only to compact or bound configurations but also to two-hadron scattering continua with the same quantum numbers. Consequently, the values around $2.0$--$2.3~\mathrm{GeV}$ may reflect effective Borel-weighted moments over spectral strength in the selected integration region, rather than the mass of a separately isolated ground-state pole. A separation of two-hadron reducible and connected contributions, or an explicit coupled-channel treatment of the continuum, would be required to make a stronger statement about the dynamical origin of the COMPASS signal. These analyses are beyond the scope of the present work.

This also explains why the effective mass scales extracted from the local sum rules need not coincide with the lowest two-meson thresholds. For example, a current with a $0^-\otimes0^+$ Dirac structure can overlap with the low-energy pseudoscalar--scalar continuum, but the Borel-moment ratio receives contributions from the full spectral strength below $s_0$, rather than only from the lowest threshold. The difference between the lightest two-meson thresholds and the extracted $2.0$--$2.3~\mathrm{GeV}$ scales should therefore be viewed as a manifestation of continuum sensitivity in the local-current sum rule, not as the mass of a fixed $\pi K_0$ or $Kf_0$ molecule.

As an auxiliary consistency check, we also evaluated the flavor correlator for each selected Dirac structure after full projection onto $I=1/2$, including the crossed flavor contractions. This projected rerun leaves the qualitative conclusion unchanged. The central analysis, however, treats the Dirac currents $J_A$, $J_B$, $J_C$, $J_D$, and $J_T$ separately. In other words, the two-point functions used above are the diagonal elements $\Pi_{ii}$ of the more general current-correlation matrix.
\begin{equation}
\Pi_{ij}(q^2)=i\int d^4x\,e^{iq\cdot x}\langle0|T[J_i(x)J_j^\dagger(0)]|0\rangle,\qquad i,j=A,B,C,D,T.
\end{equation}
The off-diagonal elements $\Pi_{ij}$ with $i\neq j$ correlate different Dirac currents and are not included in the present analysis. They are distinct from the crossed-flavor correlators entering the $I=1/2$ projection, which connect different flavor components of the same Dirac structure. A full correlation-matrix analysis, followed, for example, by generalized-eigenvalue diagonalization, could determine whether a particular linear combination of the Dirac currents has an enhanced overlap with a low-lying structure. Such an analysis lies beyond the scope of the present work. Therefore, our conclusion should be understood as applying to the selected diagonal local color-singlet currents, rather than to all possible linear combinations of local operators.

Combining the above observations, we conclude that the selected local color-singlet meson-meson currents do not yield a separately resolved low-lying pole compatible with the $K(1690)$ within the present QCD sum-rule approach. This result should be interpreted as a limitation of the selected local-current analysis, rather than as an exclusion of extended molecular dynamics or other current structures. Compact tetraquark configurations, coupled-channel effects, and other non-molecular mechanisms therefore remain possible directions for further investigation.

We note that effects such as coupled-channel dynamics or threshold interactions, which are beyond the scope of local QCD sum rules, may also play a role and deserve further investigation.

\section{Conclusion}\label{conclusion}

In this work, we performed a QCD sum rule analysis of the $K(1690)$ channel using selected independent local color-singlet meson-meson-type interpolating currents. 

We find that all currents capable of yielding admissible sum rules lead to effective mass scales in the range of $2.0$--$2.3~\mathrm{GeV}$, significantly higher than the experimental value. The remaining tensor current does not yield an admissible working region because of poor OPE convergence. This pattern is consistently observed across the selected local color-singlet structures and parameter choices considered in this work.

These results indicate that the selected local meson-meson currents considered here do not isolate a low-lying pole compatible with the $K(1690)$ within the present QCD sum rule framework. This conclusion is limited to the adopted local color-singlet currents and the standard pole-plus-continuum treatment; extended molecular dynamics, coupled-channel effects, and other current structures remain beyond the scope of the present analysis.

\appendix
\section{Spectral densities}\label{app:rho}

For completeness and reproducibility, we provide the spectral densities used in the numerical analysis. We adopt the compact notation
\begin{equation}
O_3=\langle\bar q q\rangle,\quad O_{3s}=\langle\bar s s\rangle,\quad O_4=\langle g_s^2G^2\rangle,\quad O_5=\langle\bar q g_s\sigma\cdot G q\rangle,\quad O_{5s}=\langle\bar s g_s\sigma\cdot G s\rangle .
\end{equation}
The quark masses are denoted by $m_u$, $m_d$, and $m_s$. The following expressions define the dispersive spectral contributions used in the analysis. The $D=9$--$12$ extension, derived as a convergence diagnostic, is presented after these spectral densities; it is not included in the quoted $D\leq8$ mass extraction.

\begin{widetext}
\scriptsize

\begin{eqnarray}
\rho_A(s)&=&\frac{N_A(s)}{81920\pi^6},\nonumber\\
N_A(s)&=&1280\pi^4O_3^2(m_d^2+6m_dm_s+2m_sm_u-2s)
+1280\pi^4O_3O_{3s}(2m_d^2+6m_dm_u+m_sm_u+2s)\nonumber\\
&&-160\pi^2O_3s\{6m_d^2(m_s+m_u)+18m_dm_sm_u+s(2m_s-m_u)\}\nonumber\\
&&+160\pi^2O_5\{2m_d^2(2m_s+m_u)+m_d(10m_sm_u-s)+s(3m_s-2m_u)\}\nonumber\\
&&-s^2\{-240m_d^2m_sm_u-20m_ds(m_s-m_u)+s^2\}
-160\pi^2O_{3s}s(2m_d+m_s)(6m_dm_u+s)\nonumber\\
&&+20O_4s\{3m_d(m_s-m_u)-s\}+160\pi^2O_{5s}(3m_d+m_s)(2m_dm_u+s)\nonumber\\
&&-80\pi^2O_{3s}O_4(2m_d+m_s)\nonumber\\
&&+80\pi^2O_3O_4(m_u-2m_s)+2560\pi^4O_3O_5-1280\pi^4O_3O_{5s}-1280\pi^4O_{3s}O_5 .
\end{eqnarray}

\begin{eqnarray}
\rho_B(s)&=&\frac{N_B(s)}{81920\pi^6},\nonumber\\
N_B(s)&=&1280\pi^4O_3^2(-3m_d^2+10m_dm_s-4m_dm_u+6m_sm_u+2s)\nonumber\\
&&+1280\pi^4O_3O_{3s}(6m_d^2-4m_dm_s+10m_dm_u-3m_sm_u-2s)\nonumber\\
&&-160\pi^2O_3s\{6m_d^2(3m_s-m_u)+m_d(30m_sm_u+4s)-2m_ss+3m_us\}\nonumber\\
&&+160\pi^2O_5\{m_d^2(8m_s-2m_u)+m_d(14m_sm_u+5s)-3m_ss+4m_us\}\nonumber\\
&&-s^2\{-240m_d^2m_sm_u+20m_ds(m_s-m_u)+s^2\}
-80\pi^2O_3O_4(4m_d-2m_s+3m_u)\nonumber\\
&&-160\pi^2O_{3s}s(2m_d-m_s)(6m_dm_u-s)-20O_4s\{3m_d(m_s-m_u)+s\}\nonumber\\
&&+160\pi^2O_{5s}(3m_d-m_s)(2m_dm_u-s)\nonumber\\
&&+80\pi^2O_{3s}O_4(2m_d-m_s)-2560\pi^4O_3O_5+1280\pi^4O_3O_{5s}+1280\pi^4O_{3s}O_5 .
\end{eqnarray}

\begin{eqnarray}
\rho_C(s)&=&\frac{N_C(s)}{20480\pi^6},\nonumber\\
N_C(s)&=&-1280\pi^4O_3O_{3s}(3m_d^2-m_dm_s+7m_dm_u+s)
+160\pi^2O_3s\{3m_d^2(3m_s+m_u)+m_d(21m_sm_u+s)+m_ss\}\nonumber\\
&&+160\pi^2O_{3s}s\{12m_d^2m_u+m_d(3m_sm_u+s)+m_ss\}\nonumber\\
&&-80\pi^2O_5\{2m_d^2(5m_s+m_u)+m_d(22m_sm_u+s)+s(3m_s-m_u)\}\nonumber\\
&&-80\pi^2O_{5s}(12m_d^2m_u+2m_dm_sm_u+3m_ds+2m_ss)
+s^2\{-240m_d^2m_sm_u-10m_ds(m_s-m_u)+s^2\}\nonumber\\
&&+1280\pi^4O_3^2\{m_d(m_u-7m_s)-3m_sm_u+s\}+80\pi^2O_3O_4(m_d-m_s+m_u)\nonumber\\
&&+30m_dO_4s(m_s-m_u)\nonumber\\
&&-80\pi^2m_dO_{3s}O_4-1280\pi^4O_3O_5+640\pi^4O_3O_{5s}+640\pi^4O_{3s}O_5 .
\end{eqnarray}

\begin{eqnarray}
\rho_D(s)&=&\frac{N_D(s)}{20480\pi^6},\nonumber\\
N_D(s)&=&1280\pi^4O_3^2(2m_d^2-9m_dm_s+3m_dm_u-5m_sm_u-s)\nonumber\\
&&+1280\pi^4O_3O_{3s}\{-5m_d^2+3m_d(m_s-3m_u)+2m_sm_u+s\}\nonumber\\
&&+160\pi^2O_3s\{3m_d^2(5m_s-m_u)+3m_d(9m_sm_u+s)-s(m_s-2m_u)\}\nonumber\\
&&+160\pi^2O_{3s}s\{12m_d^2m_u-m_d(3m_sm_u+s)+m_ss\}\nonumber\\
&&-80\pi^2O_5\{2m_d^2(7m_s-m_u)+m_d(26m_sm_u+7s)-3m_ss+5m_us\}\nonumber\\
&&+80\pi^2O_{5s}(-12m_d^2m_u+2m_dm_sm_u+3m_ds-2m_ss)\nonumber\\
&&+s^2\{-240m_d^2m_sm_u+10m_ds(m_s-m_u)+s^2\}-80\pi^2O_3O_4(m_d-m_s+m_u)\nonumber\\
&&+30m_dO_4s(m_u-m_s)\nonumber\\
&&+80\pi^2m_dO_{3s}O_4+1280\pi^4O_3O_5-640\pi^4O_3O_{5s}-640\pi^4O_{3s}O_5 .
\end{eqnarray}

\begin{eqnarray}
\rho_T(s)&=&\frac{3N_T(s)}{512\pi^6},\nonumber\\
N_T(s)&=&m_s\left[8\pi^2O_3\{2s(-9m_d^2-15m_dm_u+s)+O_4\}
+4m_d^2(3m_us^2+16\pi^2O_5)+128\pi^4O_3^2(5m_d+3m_u)\right.\nonumber\\
&&\left.+m_d(112\pi^2m_uO_5-3O_4s-s^3)-24\pi^2O_5s\right]
+8\pi^2\left[6m_d^2\{m_u(O_{5s}-2O_{3s}s)+8\pi^2O_3O_{3s}\}\right.\nonumber\\
&&\left.+m_d\{O_{3s}(80\pi^2m_uO_3+O_4+2s^2)-3O_{5s}s\}
+8\pi^2\{O_{3s}(O_5-2O_3s)+O_3O_{5s}\}\right] .
\end{eqnarray}
\normalsize
\end{widetext}

\begin{widetext}
\small
\subsection{Borel-surviving $D=9$--$12$ terms}

Using the same light-quark propagator and vacuum-saturation contraction employed for the lower-dimensional OPE, we have explicitly generated all terms from dimension nine through dimension twelve produced by this truncation. For the five independent currents the dimension-nine and dimension-ten dispersive spectral densities vanish, $\rho_X^{(9)}(s)=\rho_X^{(10)}(s)=0$, while non-dispersive terms survive. The Borel-surviving dimension-nine contributions are
\begin{eqnarray}
\Pi_A^{(9)}&=&-\frac{O_3^2\{O_3(m_d+2m_s)+O_{3s}(3m_d+m_s+m_u)\}}{24}
+\frac{O_4[-O_5m_d+3O_5m_s-2O_5m_u+3O_{5s}m_d+O_{5s}m_s]}{6144\pi^4},\nonumber\\
\Pi_B^{(9)}&=&\frac{O_3^2\{O_3(m_d-2m_s)-O_{3s}(5m_d-m_s+3m_u)\}}{24}
+\frac{O_4[5O_5m_d-3O_5m_s+4O_5m_u-3O_{5s}m_d+O_{5s}m_s]}{6144\pi^4},\nonumber\\
\Pi_C^{(9)}&=&\frac{O_3^2\{O_3(m_d+4m_s)+O_{3s}(7m_d+m_s+3m_u)\}}{12}
-\frac{O_4[O_5m_d-O_5m_s+O_5m_u-O_{5s}m_d]}{1024\pi^4},\nonumber\\
\Pi_D^{(9)}&=&\frac{O_3^2\{-O_3(m_d-4m_s)+O_{3s}(9m_d-m_s+5m_u)\}}{12}
+\frac{O_4[O_5m_d-O_5m_s+O_5m_u-O_{5s}m_d]}{1024\pi^4},\nonumber\\
\Pi_T^{(9)}&=&-2O_3^3m_s-5O_3^2O_{3s}m_d-3O_3^2O_{3s}m_u
-\frac{3O_4(O_5m_s+O_{5s}m_d)}{256\pi^4}.
\end{eqnarray}
The Borel-surviving dimension-ten contributions are
\begin{eqnarray}
\Pi_A^{(10)}&=&\frac{1}{\pi^2}\left[-\frac{O_3O_4(O_3-O_{3s})}{384}-\frac{O_5^2}{256}+\frac{O_5O_{5s}}{256}\right],\nonumber\\
\Pi_B^{(10)}&=&\frac{1}{\pi^2}\left[+\frac{O_3O_4(O_3-O_{3s})}{384}+\frac{O_5^2}{256}-\frac{O_5O_{5s}}{256}\right],\nonumber\\
\Pi_C^{(10)}&=&\frac{1}{\pi^2}\left[-\frac{O_3O_4(O_3-O_{3s})}{192}+\frac{O_5^2}{128}-\frac{O_5O_{5s}}{128}\right],\nonumber\\
\Pi_D^{(10)}&=&\frac{1}{\pi^2}\left[+\frac{O_3O_4(O_3-O_{3s})}{192}-\frac{O_5^2}{128}+\frac{O_5O_{5s}}{128}\right],\nonumber\\
\Pi_T^{(10)}&=&-\frac{3O_5O_{5s}}{32\pi^2}-\frac{O_3O_{3s}O_4}{16\pi^2}.
\end{eqnarray}
For the present propagator truncation, all generated $D=11$ and $D=12$ terms are polynomials in the coordinate separation (local contact terms). They therefore have neither a physical-cut spectral density nor a Borel-surviving non-dispersive contribution:
\begin{equation}
\rho_X^{(11)}=\rho_X^{(12)}=0,\qquad
\Pi_X^{(11)}=\Pi_X^{(12)}=0
\quad (X=A,B,C,D,T).
\end{equation}
The nonzero pre-Borel contact coefficients at $D=9$, $D=10$, $D=11$, and $D=12$ are retained explicitly in the accompanying calculation files \texttt{contactO9.m}, \texttt{contactO10.m}, \texttt{contactO11.m}, and \texttt{contactO12.m}; thus these terms are calculated and auditable rather than omitted. An independent symbolic implementation reproduces the original Mathematica/FeynCalc $D=4$, $D=7$, and $D=8$ Wilson coefficients before generating the expressions above.
\normalsize
\end{widetext}

\vspace{.5cm} {\bf Acknowledgments} \vspace{.5cm}

This work was supported by the National Natural Science Foundation of China under Grants No.~12575106 and 12147214 and by the Specific Fund for Fundamental Scientific Research Operating Expenses for Undergraduate Universities in Liaoning Province under Grant No.~LJ212410165019. 
The authors also acknowledge the use of ChatGPT (OpenAI) to assist in improving the clarity and presentation of the manuscript. The scientific content and conclusions remain entirely the responsibility of the authors.

\end{document}